\documentclass[manuscript]{aastex63}
\usepackage{CJK}

\shorttitle{An overdensity of red galaxies around W1835$+$4355 }
\shortauthors{Luo et al.}

\begin{document}
\begin{CJK*}{UTF8}{gbsn}

\title{An overdensity of red galaxies around the hyperluminous dust-obscured quasar W1835$+$4355 at $z=2.3$ }

\correspondingauthor{Yibin Luo,Lulu Fan}
\email{yibinluo@mail.ustc.edu.cn,llfan@ustc.edu.cn}

\author[0000-0002-8079-6525]{Yibin Luo （罗毅彬）}
\affiliation{CAS Key Laboratory for Research in Galaxies and Cosmology, Department of Astronomy, University of Science and Technology of China, Hefei 230026, China}

\affil{School of Astronomy and Space Sciences, University of Science and Technology of China, Hefei, Anhui 230026, People's Republic of China}

\author[0000-0003-4200-4432]{Lulu Fan （范璐璐）}
\affiliation{CAS Key Laboratory for Research in Galaxies and Cosmology, Department of Astronomy, University of Science and Technology of China, Hefei 230026, China}

\affil{Institute of Deep Space Sciences, Deep Space Exploration Laboratory, Hefei 230026, China}

\affil{School of Astronomy and Space Sciences, University of Science and Technology of China, Hefei, Anhui 230026, People's Republic of China}

\author[0000-0002-6684-3997]{Hu Zou （邹虎）}
\affil{National Astronomical Observatories, Chinese Academy of Sciences, Beijing, 100012, China}

\author[0000-0001-9495-7759]{Lu Shen （沈璐）}
\affiliation{CAS Key Laboratory for Research in Galaxies and Cosmology, Department of Astronomy, University of Science and Technology of China, Hefei 230026, China}

\affil{School of Astronomy and Space Sciences, University of Science and Technology of China, Hefei, Anhui 230026, People's Republic of China}

\author[0000-0001-8078-3428]{Zesen Lin （林泽森）}
\affiliation{CAS Key Laboratory for Research in Galaxies and Cosmology, Department of Astronomy, University of Science and Technology of China, Hefei 230026, China}

\affil{School of Astronomy and Space Sciences, University of Science and Technology of China, Hefei, Anhui 230026, People's Republic of China}

\author[0000-0003-3424-3230]{Weida Hu （胡维达）}
\affiliation{CAS Key Laboratory for Research in Galaxies and Cosmology, Department of Astronomy, University of Science and Technology of China, Hefei 230026, China}

\affil{School of Astronomy and Space Sciences, University of Science and Technology of China, Hefei, Anhui 230026, People's Republic of China}

\author[0000-0003-4959-1625]{Zheyu Lin （林哲宇）}
\affiliation{CAS Key Laboratory for Research in Galaxies and Cosmology, Department of Astronomy, University of Science and Technology of China, Hefei 230026, China}

\affil{School of Astronomy and Space Sciences, University of Science and Technology of China, Hefei, Anhui 230026, People's Republic of China}

\author{Bojun Tao （陶柏钧）}
\affiliation{CAS Key Laboratory for Research in Galaxies and Cosmology, Department of Astronomy, University of Science and Technology of China, Hefei 230026, China}

\affil{School of Astronomy and Space Sciences, University of Science and Technology of China, Hefei, Anhui 230026, People's Republic of China}

\author[0000-0002-4742-8800]{Guangwen Chen （陈广文）}
\affiliation{CAS Key Laboratory for Research in Galaxies and Cosmology, Department of Astronomy, University of Science and Technology of China, Hefei 230026, China}

\affil{School of Astronomy and Space Sciences, University of Science and Technology of China, Hefei, Anhui 230026, People's Republic of China}

\begin{abstract}

\emph{Wide-field Infrared Survey Explorer} all-sky survey has discovered a new population of hot dust-obscured galaxies (Hot DOGs), which has been confirmed to be dusty quasars. Previous statistical studies have found significant overdensities of sub-millimeter and mid-IR selected galaxies around Hot DOGs, indicating they may reside in dense regions. Here we present the near-infrared ($J$ and $K_s$ bands) observations over a $7.5\arcmin \times 7.5\arcmin$ field centered on a Hot DOG W1835$+$4355 at $z \sim 2.3$ using the wide-field infrared camera on the Palomar 200-inch telescope. We use the color criterion $J-K_s>2.3$ for objects with $K_s<20$, to select Distant Red Galaxies (DRGs). We find a significant excess of number density of DRGs in W1835$+$4355 field compared to three control fields, by a factor of about 2. The overdensity of red galaxies around W1835$+$4355 are consistent with the multi-wavelength environment of Hot DOGs, suggesting that Hot DOGs may be a good tracer for dense regions at high redshift. We find that W1835$+$4355 do not reside in the densest region of the dense environment traced by itself. A possible scenario is that W1835$+$4355 is undergoing merging process, which lowers the local number density of galaxies in its surrounding region.

\end{abstract}



\section{Introduction} \label{sec:intro}

Protoclusters are progenitors of present-day massive clusters. The study of protoclusters can place constraints on cosmological simulation \citep{chiang2013,muldrew2015}. Moreover, protoclusters are excellent laboratories to study the influence of the environment on the formation and evolution of galaxies at high redshift \citep{chiang2017}. However, it is very difficult and inefficient to directly search for protoclusters at high redshift. It is necessary to find specific sources which can trace dense regions. Luminous high-redshift radio galaxies are believed to be the ancestors of local brightest cluster galaxies \citep{miley2008,collet2015}. They have been successfully used as probes of high-redshift protoclusters \citep{venemans2002,venemans2005,venemans2007,hatch2011,mawatari2012,hayashi2012,wylezalek2013,cooke2014}.

Intensive studies focused on unobscured quasars at high redshift in order to investigate whether they can also trace dense regions. However, those studies did not reach a consistent conclusion. Some studies found that quasars are in dense environments \citep{kashikawa2007,stevens2010,falder2011,trainor2012,husband2013,morselli2014,hennawi2015,garcia-vergara2019}, while others found that the environments of quasars are not overdense \citep{mazzucchelli2017,kikuta2017,uchiyama2019,yoon2019,uchiyama2020}.

Contrary to the inconsistent conclusions of the environmental studies of unobscured quasars at high redshift, luminous dusty galaxies have been found to be strong clustering and have been suggested to reside in dense environments \citep{scott2006,brodwin2008,chapman2009,cooray2010,hickox2012}. According to supermassive black holes (SMBHs) and their host galaxies co-evolution scenario \citep{hopkins2008,alexander2012}, these dusty galaxies represent a starburst-dominated or starburst-obscured AGN composite phase.

Based on ``W1W2-dropout" method, a new population of hyperluminous, hot dust-obscured galaxies were discovered using the \emph{Wide-field Infrared Survey Explorer} (WISE) and were called as Hot DOGs \citep{eisenhardt2012,wu2012}.  These galaxies are well detected in the 12 $\mu m$ (W3) and/or 22 $\mu m$ (W4) bands, but are very faint or even undetected in the 3.4 $\mu m$ (W1) and 4.6 $\mu m$ (W2) bands. Follow-up multiwavelength studies have revealed that Hot DOGs are extremely luminous, merger-driven, heavily dust-obscured quasars at high redshift \citep{piconcelli2015,tsai2015,assef2016,fan2016a,fan2016b,fan2017,fan2018b,fan2020,ricci2017,zappacosta2018}. Previous statistical studies have found significant overdensities of mid-IR-selected and submm-selected galaxies around Hot DOGs, indicating Hot DOGs may reside in dense regions \citep{jones2014,fan2017,assef2015}. 

In this work, we use Wide-field infrared camera (WIRC) equipped on Palomar 200-inch telescope (P200)  to study the environment of a Hot DOG, W1835$+$4355 (hereafter, W1835) at $z = 2.298$. \citet{fan2016b} performed the SED decomposition of this source, and found that about 4/5 of its total luminosity attribute to the obscured quasar activity and the rest is from starburst. Moreover, based on XMM-Newton X-ray observations, \citet{piconcelli2015} found that W1835 has a reflection-dominated spectrum due to Compton-thick absorption. Since overdensities of NIR-selected red galaxies have been shown to be good tracers of potential massive structures at $z > 2$ \citep{kajisawa2006,kodama2007,uchimoto2012}, we perform a study on environment of W1835 by quantifying the density of its surrounding red galaxies.

The paper is structured as follows. In Section \ref{sec:obs}, we present the observation and data reduction. In Section \ref{sec:selection}, we describe the color criterion. Results and discussions are presented in Section \ref{sec:results} and Section \ref{sec:discussion}, respectively. We give a brief summary in Section \ref{sec:summary}. Throughout this work, we use the Vega magnitude system and assume a flat ${\rm \Lambda}$CDM cosmology \citep{komatsu2011}, with $H_0=\rm{70\ km\ s^{-1}\ Mpc^{-1}}$, $\Omega_M = 0.3$ and $\Omega_\Lambda = 0.7$.

\section{Observation and Data Reduction}\label{sec:obs}

The $J$ and $K_s$-band imaging of W1835 field was observed with P200 WIRC. This camera is mounted at the prime focus of P200, providing a field-of-view (FOV) of $\sim$ 8.7$\times$8.7 arcmin$^2$. The WIRC was equipped with a 2048$\times$2048 Hawaii-II HgCdTe detector with a pixel scale of about 0\arcsec.25. The observations took place on 2017 June 2. The total exposure times for $J$ and $K_s$-bands are about 5.0 and 1.5 hours, respectively.

\begin{deluxetable*}{ccccc}
\tablecaption{Summary of the observational data
\label{table:obserinfo}}
\tablehead{
\colhead{Name} & \colhead{Redshift} & \colhead{FOV} 
& \colhead{Exp.time(hr)} & \colhead{PSF \ FWHM($\arcsec$)}  \\
\colhead{} & \colhead{} & \colhead{(arcmin$^2$)} & \colhead{$J$ \ $K_s$} & \colhead{$J$ \ $K_s$}  }
\startdata
W1835 + 4355 & 2.298 & 56.25 & 5.0 \ 1.5 & 1.25 \ 1.02 \\
\enddata
\end{deluxetable*}

The raw data were preprocessed with overscan subtracted. The estimated dark current is about 0.26 e$^-$/s and corrected for each science exposure. The ``super" sky flats are constructed by combining all-night science frames and are then used for flat-fielding. The images after the flat-field correction are calibrated with the reference catalog of the Two Micron All-Sky Survey (2MASS). The astrometric accuracy relative to the 2MASS catalog is about 0\farcs13 for $J$ band and 0\farcs19 for $K_s$ band. The flux is calibrated in the Vega magnitude system. The calibrated frames are stacked to generate a deeper image for each band by using \textsc{SWarp} \citep{bertin2002}. Before stacking, those frames with large seeing and bright sky background are removed. Final stacked images have the size of $1800\times1800$ pixels$^2$ and a linear pixel scale of 0\farcs25, which provides an angular size of about 7.5 arcmin. Table \ref{table:obserinfo} gives a summary of the observations.

\begin{figure*}
\plotone{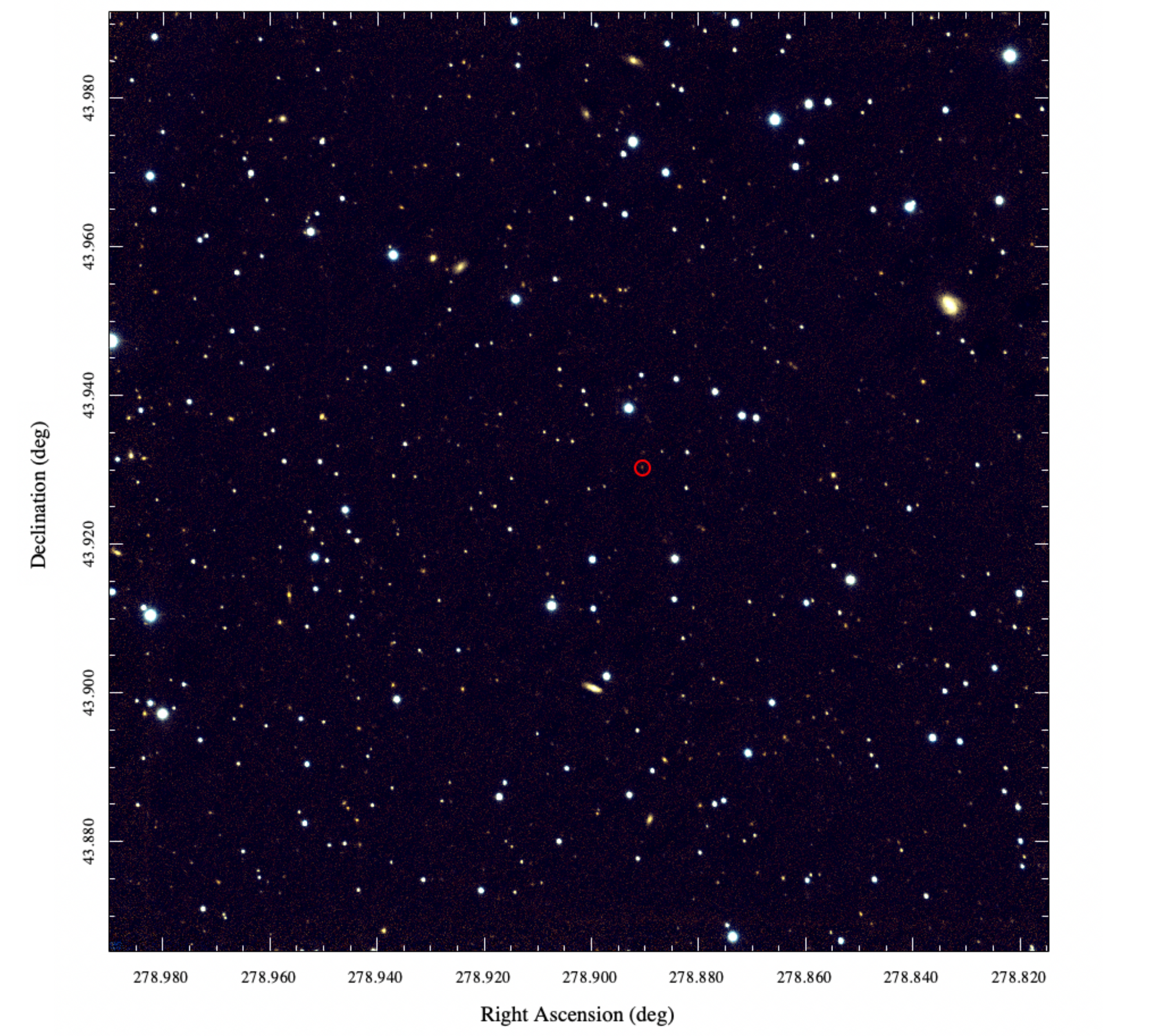}
\caption{Composited color image of the W1835 field. The image size is $7.5 \times 7.5$ arcmin$^2$. The red open circle marks the position of W1835.
\label{fig-rgb}}
\end{figure*}

The photometric catalog was generated from the stacked images. The FWHMs of the point spread function (PSF) are 1\farcs25  and 1\farcs02 in the $J$ and $K_s$ band, respectively. In order to measure the colors within the same aperture across two bands, the $K_s$ band image was convolved with a kernel to match the PSF of $J$ band image taken under the worst seeing. The PSF match was performed using \textsc{photutils} package in \textsc{Python} \citep{Bradley2020}. Photometry was performed by running \textsc{SExtractor} \citep{bertin1996} in dual-image mode. The $K_s$ band image was used for the source detection, while the $J$ band image and the convolved $K_s$ band image were used for photometry. For the color measurements in $J - K_s$, we used aperture magnitude MAG\_APER with $2\farcs5$ aperture diameter. We adopted MAG\_AUTO as the total $K_s$ band magnitude of detected objects. The $5\sigma$ limiting magnitudes of the $J$ and $K_s$ bands are 21.8 and 20.0 mag, respectively. Our source detection completeness in the $K_s$ band is 90\% at $K_s = 19.4$ mag and 50\% at $K_s = 19.9$ mag. Galactic extinction is estimated at the position of W1835, based on \citet{schlafly2011}\footnote{\url{https://irsa.ipac.caltech.edu/applications/DUST/}}. All the magnitudes in the field are corrected for Galactic extinction. Figure \ref{fig-rgb} shows the composited color image of the W1835 field, where the blue, red, and green channels are $J$, $K_s$ and and the average of these two images, respectively.

\section{COLOR SELECTION CRITERION}\label{sec:selection}

Considering the rest-frame optical spectral features of red galaxies dominated by evolved stars, we can use the Balmer/4000\AA\ break to select those galaxies. The Balmer break at 3646\AA\ is due to electrons bound-free absorption, which is strongest in A class stars. The 4000\AA\ break is due to absorption from ionized metals, which occurs in older stars \citep{hammer2017}. Since these breaks enter between $J$ and $K_s$ bands at $2 \leq z \leq 4$, a red $J - K_s$ color can be an effective criterion to select high-redshift galaxies. Following the criterion from \citet{franx2003}, we select objects with $J-K_s > 2.3$ as distant red galaxies (DRGs). These galaxies may represent the progenitors of massive galaxies at the present day. Many DRGs have had follow-up spectroscopic observations and have been confirmed to be at $2 < z < 3$ \citep{vandokkum2003,schreiber2004,reddy2005}.

\section{RESULTS}\label{sec:results}

\subsection{Color-Magnitude Diagram}\label{sec:CMD}

\begin{figure}
\plotone{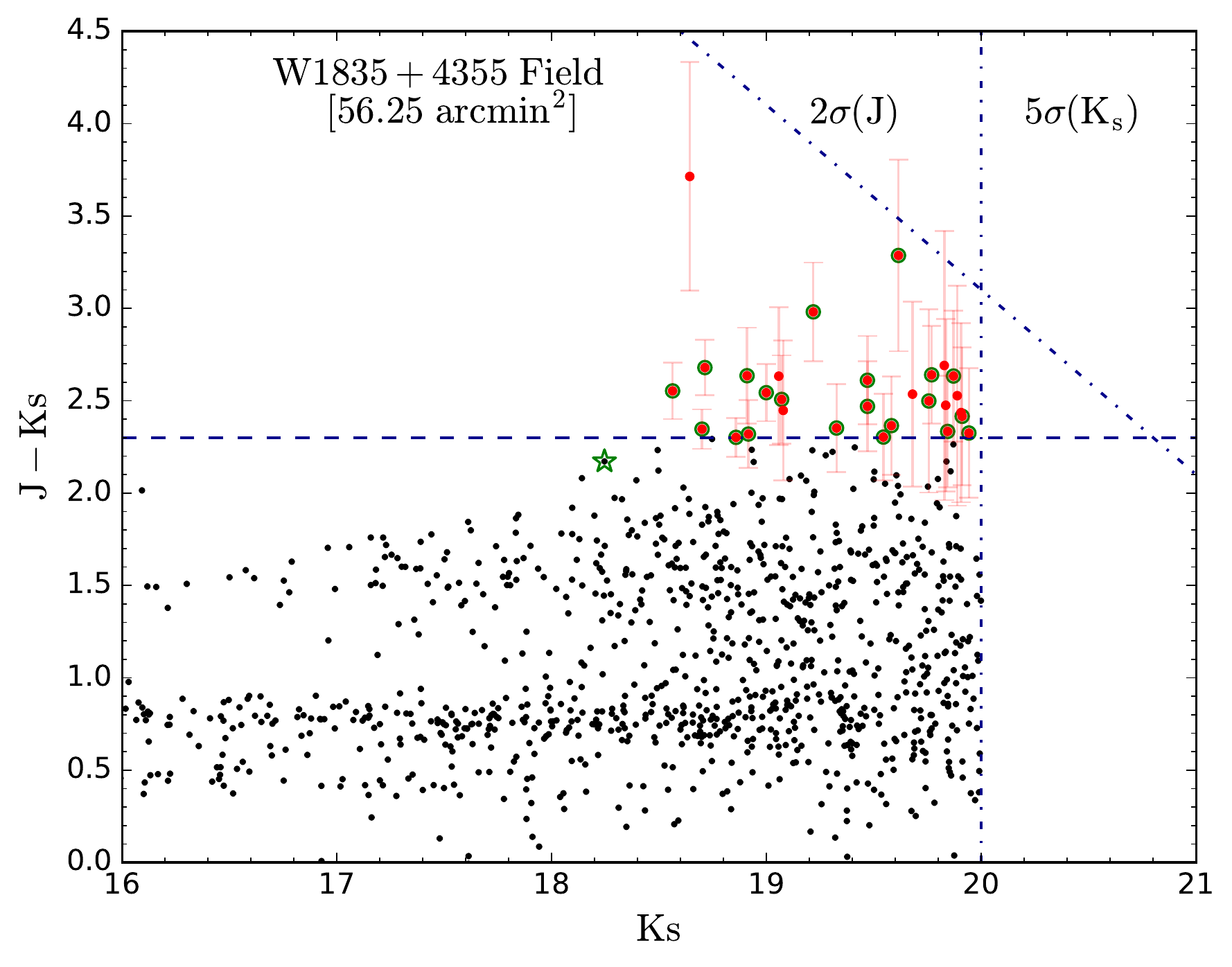}
\caption{
Color-magnitude diagram of sources in the W1835 field. The $2\sigma$ limiting magnitudes of $J$-band and the $5\sigma$ limiting magnitudes of $K_s$-band are indicated by the dash-dotted lines. The horizontal dashed line corresponds to $J - K_s = 2.3$. Sources above this line are selected as DRGs and are marked as red filled circles, while the others are shown as the black dots. Those DRGs confirmed by visual examination are indicated by green circles. The open star marked the Hot DOG W1835. The error bars show the photometric errors for color measurements.
\label{fig:W1835 + 4355 field}}
\end{figure}

We use the $2\sigma$ limiting magnitude of the $J$-band ($J=22.9$) to select DRGs down to the $5\sigma$ limiting magnitude of $K_s$-band ($K_s=20.0$), which is commonly used in previous studies of DRG selection \citep{kajisawa2006,kodama2007}. Figure \ref{fig:W1835 + 4355 field} shows a $K_s$ vs $J - K_s$ color-magnitude diagram of sources in the W1835 field. DRGs are plotted as red filled circles. 29 sources are selected by using the color selection criterion. We perform a visual inspection of these sources to rule out possible spurious sources. We visually confirmed 21 sources, represented by green circles in Figure \ref{fig:W1835 + 4355 field}. Only the visually confirmed sources are used for the following analyses. We notice that although our selected DRGs are not a pure sample as many of them have $J - K_s$ color close to the selection limit ($J-K_s = 2.3$), this effect has been taken into account in the forthcoming results. The details can be found in Section \ref{sec:Cumulative}.

\subsection{Control Fields}\label{sec:control}

\begin{deluxetable}{cccc}
\tablecaption{Control Fields 
\label{table:controlinfo}}
\tablehead{
\colhead{Field} & \colhead{Survey} & \colhead{Area} & \colhead{$5\sigma$ Limiting magnitudes} \\
\colhead{} & \colhead{} & \colhead{} & \colhead{$J$ \ \ \ $K_s$} }
\startdata
COSMOS    &       UltraVISTA      &   1.62 \ deg$^2$ &  23.4 \ \ 22.0  \\
GOODS-N       &       MODS        &       103 \ arcmin$^2$ &  23.9 \ \ 22.8  \\
GOODS-S       &       ESO/GOODS        &       173 \ arcmin$^2$ &  24.0 \ \ 22.5  \\
\enddata

\end{deluxetable}

We use the deep NIR data from three surveys: UltraVISTA  in the Cosmic Evolution Survey (COSMOS) field \citep{mccracken2012,muzzin2013}, MOIRCS Deep Survey (MODS) in the Great Observatories Origins Survey Northern (GOODS-N) \citep{kajisawa2011}  and ESO/GOODS in Southern (GOODS-S) field \citep{retzlaff2010}. All these fields have $K_s$-selected NIR color catalogs. The depths of the $K_s$-band of three control fields are much deeper than that of the W1835 field. The completeness of three control fields at $K_s = 20$ is 100\%. The details of each field are listed in Table \ref{table:controlinfo}.

\begin{figure*}
\plotone{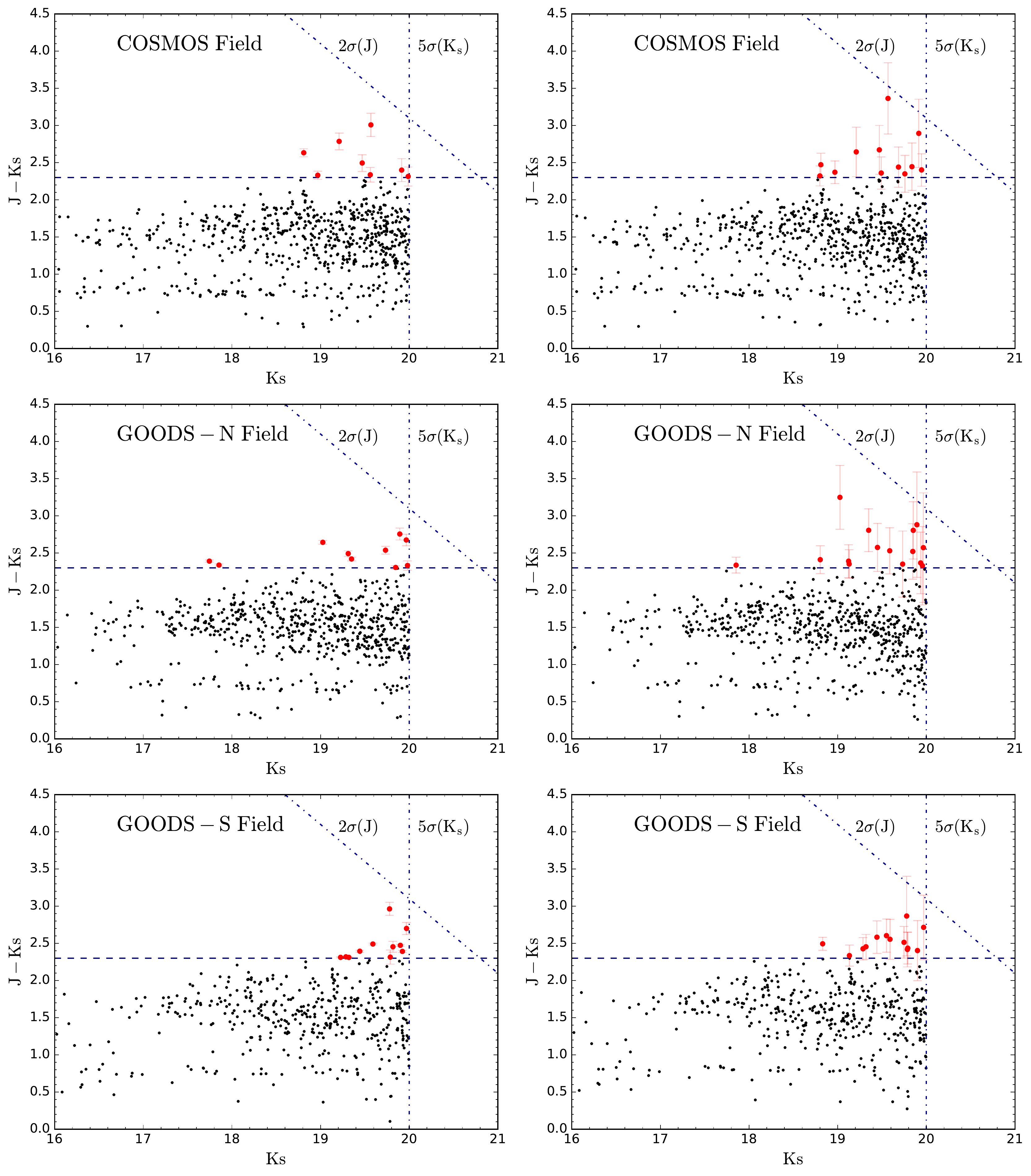}
\caption{
Color-magnitude diagrams of three control fields COSMOS, GOODS-N and GOODS-S. We randomly select a 56.25 arcmin$^2$ area in each control field. Symbols are the same as in Figure \ref{fig:W1835 + 4355 field}. The raw color-magnitude diagrams of three control fields from their catalogs are plotted in the left panel. We plot the matched color-magnitude diagrams of three control fields in the right panel. The magnitudes of the sources in the right panel are scattered based on the matched magnitude errors, as a Monte Carlo simulation example.
\label{fig:control field}}
\end{figure*}

We directly use the catalogs of the three control fields and select DRGs using the same color criteria and limiting magnitudes obtained in the W1835 field. We plot color-magnitude diagrams of the three control fields, respectively. The areas of the three control fields are larger than that of the W1835 field. In order to have a direct comparison with the W1835 field, we randomly selected an area of 56.25 arcmin$^2$ in each control field and select DRGs in these areas. For each control field, this step is performed several times, resulting in a distribution of the number of DRGs which is found to approximately follow the Poisson distribution. The median numbers of DRGs are 8, 10, 11 for COSMOS, GOODS-N and GOODS-S, respectively, which are significantly less than that in the W1835 field. In the left panel of Figure \ref{fig:control field}, we plot the raw color-magnitude diagrams of three control fields from their catalogs.

We notice that three control fields all show smaller magnitude errors compared to the W1835 field data at any given magnitude. The photometric error has effect on the color selection. The sources may scatter into or out of the selection region, causing contamination or missing sources in DRG selection. This effect could be different when the magnitude error is different. To make the W1835 field and control fields data share the same effect, we match the noise of the control fields to the W1835 field data. We fit the magnitude-magnitude error distribution of the W1835 field data and match the magnitude of the control fields to the corresponding magnitude error based on the fitting curve. We plot the matched color-magnitude diagrams of three control fields in the right panel of Figure \ref{fig:control field}. To directly show the effect of photometric error on color selection, the magnitudes of the sources in the right panel are scattered based on the matched magnitude errors. The detail of analyzing the effect of photometric error on color selection results is described in Section \ref{sec:Cumulative}. We show the result of a single Monte Carlo simulation in this panel as an example.

\subsection{Cumulative Number Density of DRGs}\label{sec:Cumulative}

In order to quantitatively describe whether there is overdensity in the red galaxies around W1835, we show the cumulative number densities of DRGs as a function of $K_s$ magnitude in the W1835 field and three control fields. The number density is obtained by dividing the number $N$ by the area $S$. $\sqrt{N}/S$ represents the uncertainty of the number density based on Poisson statistics.

\begin{figure*}
\plotone{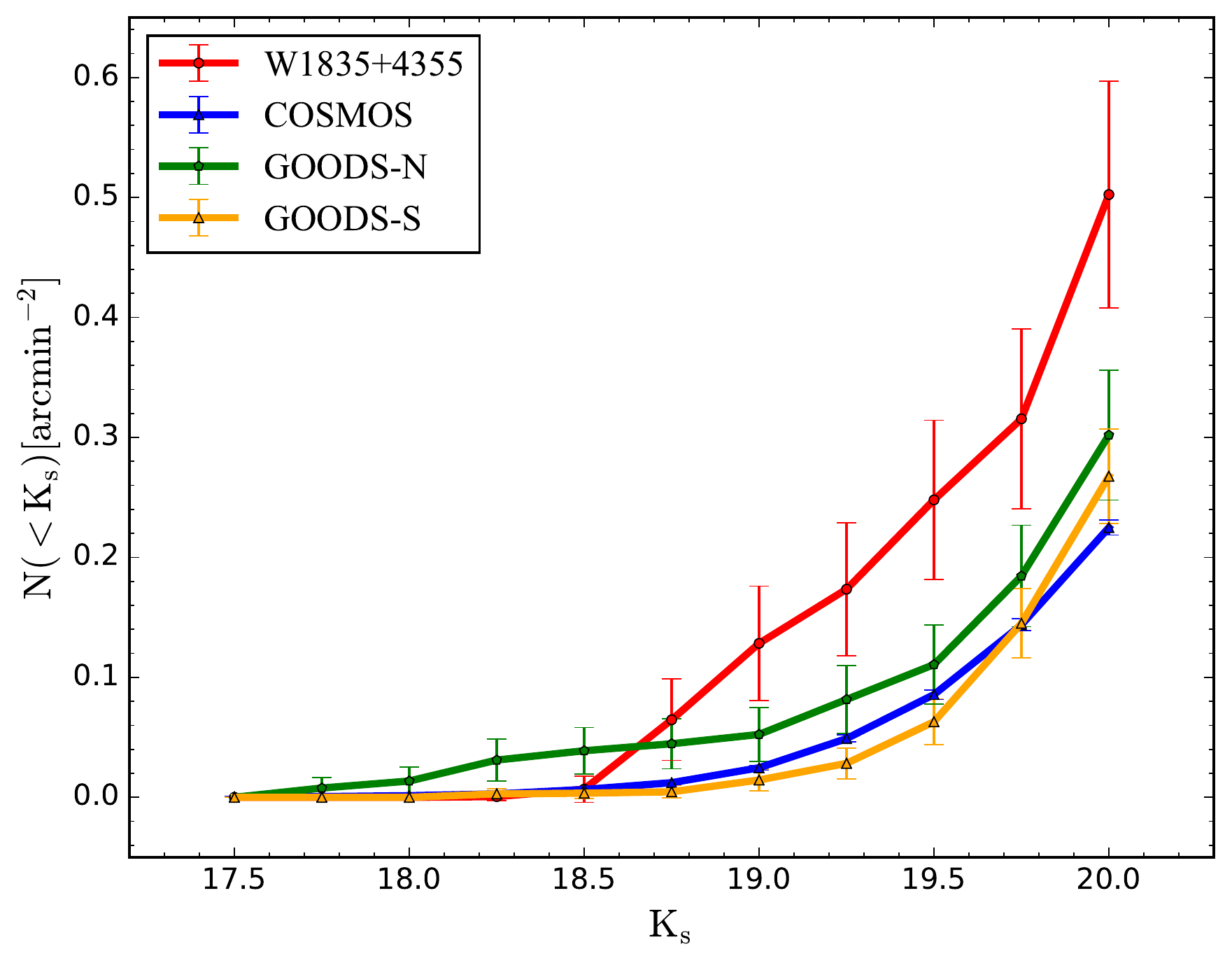}
\caption{
The cumulative number densities of DRGs in the W1835 field and three control fields as a function of $K_s$ magnitudes. Error bars are based on Poisson statistics. We show the results of the Monte Carlo simulation, with the W1835 field data and the full areas of three control fields with noise matching. Notice that the W1835 field data have corrected for the completeness.
\label{fig:cum}}
\end{figure*}

We use the W1835 field data and the noise-matched data of randomly selecting a 56.25 arcmin$^2$ area in each control field, which described in Section \ref{sec:control}, to analyze the effect of photometric error on color selection results. We utilize Monte Carlo simulation to analyze the effect of photometric error on the results. For each source, we randomly assign mock $J$ and $K_s$ magnitude based on its measured magnitude and associated photometric error.
The mock magnitude is randomly selected from a Gaussian distribution with the measured magnitude as mean value and its associated error as standard deviation. We repeat this process for 1000 times, and calculate the probability of each source of being selected as a DRG. We add the probabilities of all sources within the limiting magnitude and divide it by the area to derive the number density. With Monte Carlo simulation, the total overdensity factors of W1835 field compared to COSMOS, GOODS-N and GOODS-S fields without completeness correction are $1.76 \pm 0.63 $, $1.34 \pm 0.44 $, and $1.65 \pm 0.58 $, respectively. After completeness correction, the factors mentioned above are corrected to $2.30 \pm 0.78 $, $1.76 \pm 0.55 $, and $2.16 \pm 0.72 $, respectively. We also do the Monte Carlo simulation for the full areas of three control fields, as listed in Table \ref{table:controlinfo}. We show the cumulative number densities of DRGs in the Figure \ref{fig:cum}. The total overdensity factors of W1835 field compared to COSMOS, GOODS-N and GOODS-S fields after completeness correction are $2.23 \pm 0.42 $, $1.66 \pm 0.43 $, and $1.88 \pm 0.45 $, respectively. The results for the full areas of three control fields are consistent with the results from randomly selecting a 56.25 arcmin$^2$ area in each control field. The average total overdensity factor of W1835 field is about 2. Due to the limitation of observation depth, only the bright DRGs can be detected in this work. Previous studies in protoclusters have found overdensity factor of DRGs similar to our results \citep{kajisawa2006,kodama2007}, but their DRGs reach down to a much fainter magnitude limit. Therefore, we suggest that the W1835 field has a significant overdensity of the bright red galaxies compared to the three control fields. We expect this overdensity to remain true for the fainter red galaxies, but this will require future observations at deeper depths to determine.

\section{DISCUSSION}\label{sec:discussion}

\subsection{Environments of Hot DOGs}

Previous studies on environments of Hot DOGs have suggested that they may live in dense regions \citep{jones2014,assef2015,fan2017}. For instance, overdensities of sub-millimeter galaxies (SMGs) in the vicinity of Hot DOGs have been found in two small samples of Hot DOGs using JCMT SCUBA2 850 $\mu m$ observations \citep{jones2014,fan2017}. \citet{jones2014} detected one SMG within 1.5 arcmin radius of W1835, which revealed a moderate overdensity of SMGs compared with blank field survey \citep{weiss2009}. \citet{assef2015} found an overdensity of mid-IR Spitzer-selected galaxies around Hot DOGs within 1 arcmin compared with random pointing. In this work, we find an overdensity of red galaxies around W1835 compared to three control fields. Our results are consistent with overdensities of sub-millimeter and mid-IR Spitzer-selected galaxies around Hot DOGs \citep{jones2014,fan2017,assef2015}, suggesting that Hot DOGs may be a good tracer for dense regions such as protoclusters. In addition, \citet{bridge2013} found that Hot DOGs show a substantial overlap with the WISE-selected Ly$\alpha$ blobs (LABs) at $z > 2$, and W1835 is one of them. High-redshift LABs are often found in the dense environment of star-forming galaxies (SFGs) \citep{steidel2000,matsuda2011,saito2015}. W1835, as one of high-redshift LABs, is very likely lying in the dense region traced by SFGs.

\begin{figure*}
\plotone{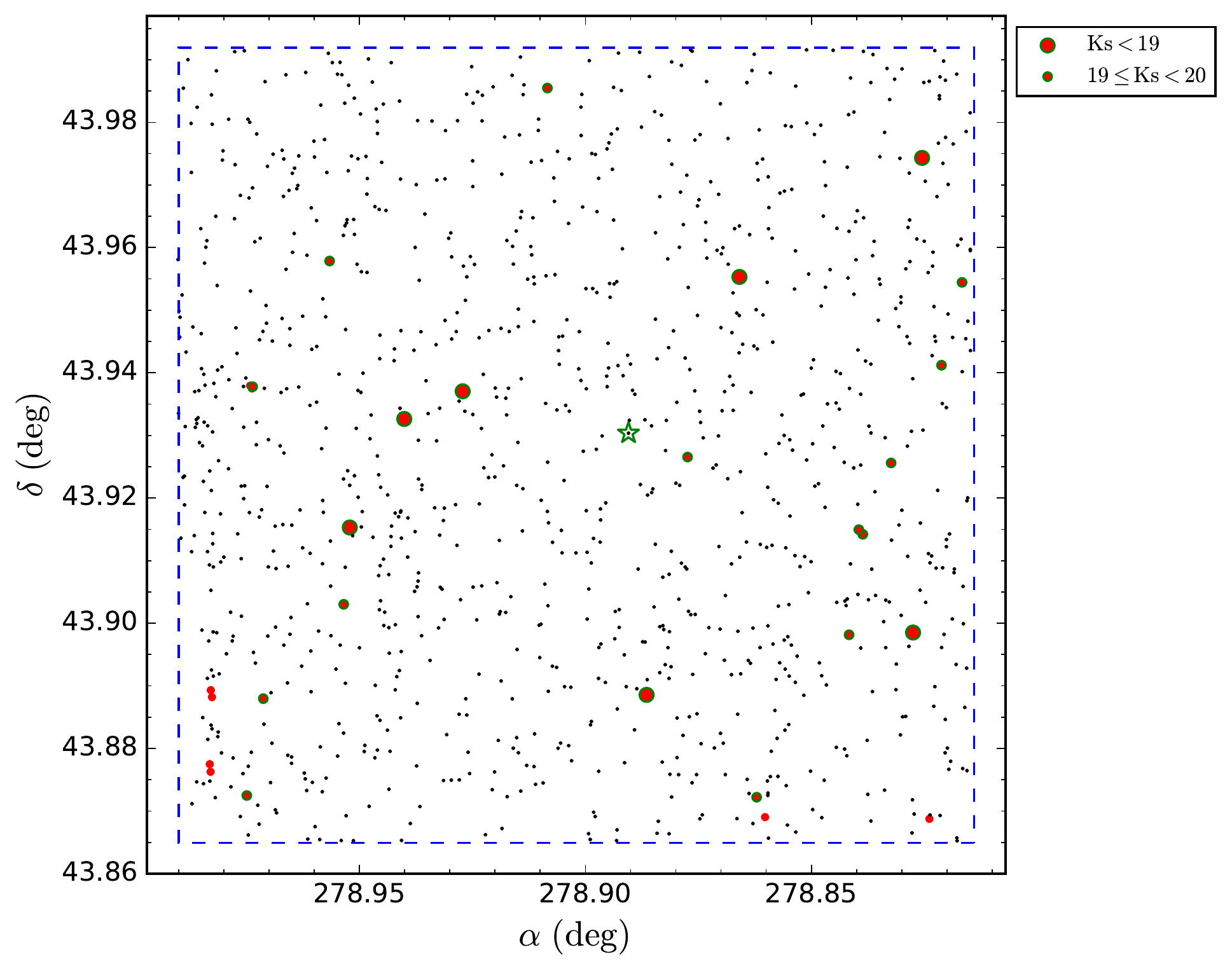}
\caption{
The spatial distribution of DRGs in W1835 field. The blue dashed box represents the $7.5\arcmin \times 7.5\arcmin$ FOV of this observation. Symbols are the same as in Figure \ref{fig:W1835 + 4355 field}.
\label{fig:spatial}}
\end{figure*}

\begin{figure}
\plotone{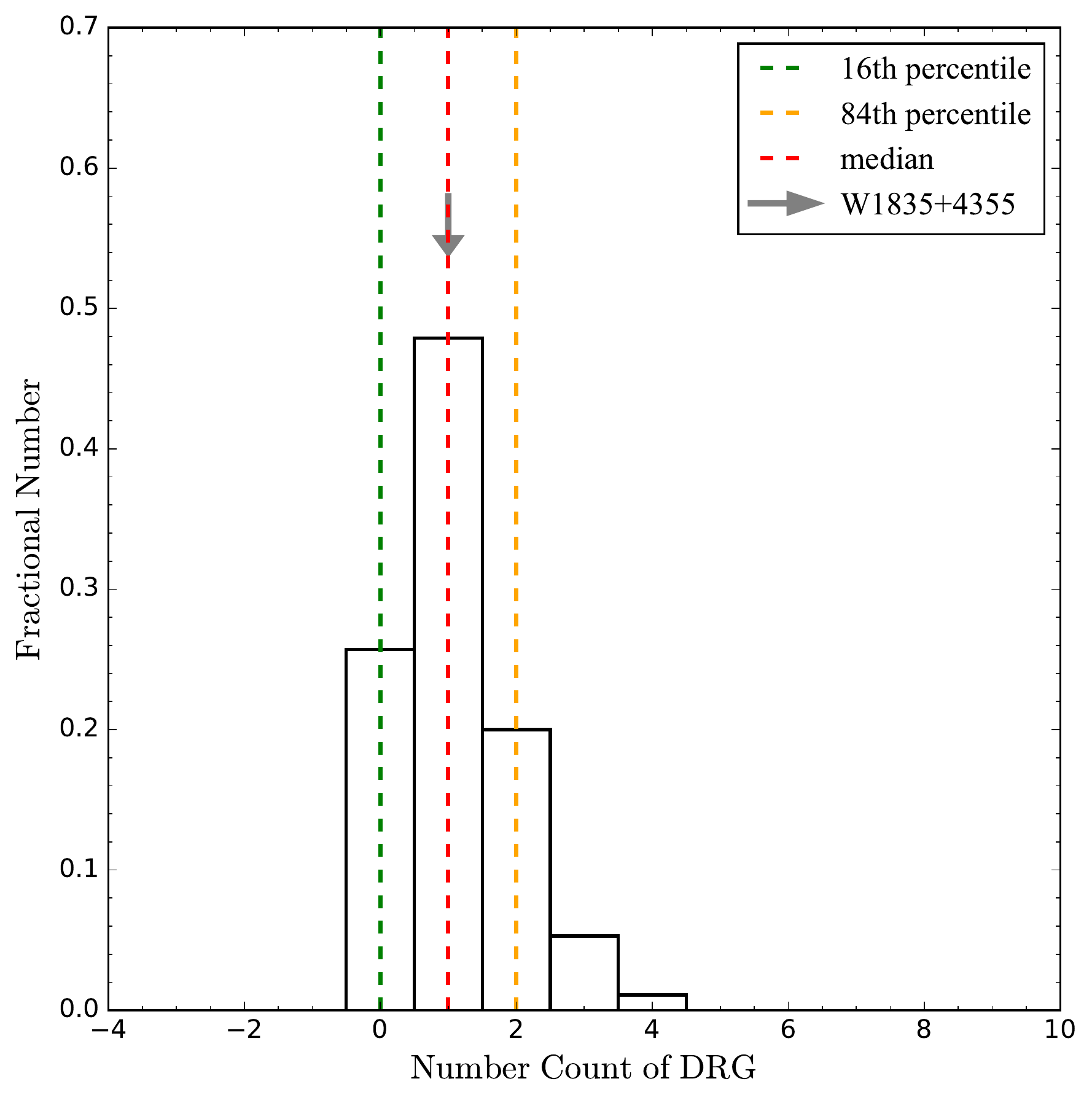}
\caption{
The distribution of the number count of DRGs in random pointings in W1835 field. The 16\% and 84\% quantiles represent the spread of the distribution, which are 0 and 2 respectively. The value calculated with W1835 as the center is marked by gray arrow.
\label{fig:local}}
\end{figure}

Furthermore, \citet{assef2015} had found that the density of mid-IR Spitzer-selected galaxies in the Hot DOG fields showed good agreement with that in the radio-loud AGN fields from CARLA survey. In addition, the overdensity of SMGs \citep{breuck2004,dannerbauer2014,rigby2014} and DRGs \citep{kajisawa2006,kodama2007} have also been found in the radio-loud AGN fields. These results suggest that the environments of Hot DOGs and radio-loud AGN may be similar.

\subsection{Local overdensity of DRGs}

We show the spatial distribution of DRGs in the W1835 field in Figure \ref{fig:spatial}. We find that W1835 does not seem to be in the most locally overdense regions. We count the number of DRGs within a fixed radius at random pointings in the W1835 field to study the local environment. Previous  studies have found that there is no hint of angular clustering of SMGs and mid-IR Spitzer-selected galaxies around Hot DOGs within $1.5\arcmin$ and $1\arcmin$, respectively \citep{jones2014,jones2017,assef2015}. We choose $1\arcmin$ as the radius, which corresponds 1.6 Mpc comoving distance at $z = 2.3$. This scale is similar to those works mentioned above. We randomly generate 1000 pointings inside the borders of the central $5.5\arcmin \times 5.5\arcmin$ region and then count the number of DRGs within $1\arcmin$ for each pointing. We show the distribution of the number count of DRGs in Figure \ref{fig:local}. The value calculated with W1835 as the center, which is 1, is marked by gray arrow. We find that this value is exactly the median value of the distribution. It indicates that the local environment of W1835 does not appear to be much denser in the field. In other words, W1835 does not reside in the densest region of the overdense environment traced by itself. This result is consistent with the idea that Hot DOGs may not be the central galaxies of the overdensities traced by themselves \citep{assef2015}. It is worth noting that \citet{shen2021} also found similar results in the proto-structure at $z = 3.3$ traced by radio-detected AGN. They proposed a scenario where merging might already have happened in this case and lowered the local density of their surrounding region. Previous studies have found X-ray-selected or IR-selected obscured AGNs have a high merger fraction compared to unobscured AGNs \citep{lackner2014,kocevski2015, donley2018}. \citet{fan2016a} analyzed the host morphology of 18 Hot DOGs using Hubble Space Telescope (HST) WFC3 imaging and found that Hot DOGs have a high merger fraction. As for W1835, high-resolution HST/WFC3 imaging shows a clear merging feature \citep{fan2016a}. It is consistent with the scenario proposed by \citet{shen2021}, which suggests that the merger scenario may also be an explanation for the low local number density of galaxies around W1835.

\section{Summary and conclusion}\label{sec:summary}

In this work, we use P200 WIRC NIR images to study the environments of a Hot DOG W1835. We use DRGs as a prior for high-redshift red galaxies based on NIR color criterion $J-K_s>2.3$ and obtain the number density of DRGs around W1835. By comparing with three control fields (COSMOS, GOODS-N, GOODS-S), we find that the number densities of DRGs in W1835 field are higher than those of three control fields. There is an overdensity of red galaxies around W1835. Our result is consistent with overdensities of SMGs and mid-IR selected galaxies around Hot DOGs found by previous statistical studies. It indicates that the environments of Hot DOGs are overdense, and Hot DOGs may be a good tracer for dense regions such as protoclusters. Furthermore, multiwavelength studies on the environments of Hot DOGs also indicate that their environments are similar to high-redshift radio-loud AGN.

We also find that W1835 are not in the locally densest part of the field. This is mainly because W1835 is in the process of merging, which lowers the galaxy density of the surrounding region. Previous SMGs and mid-IR observations of Hot DOGs' environments also find that there is no hint of angular clustering of SMGs and mid-IR Spitzer-selected galaxies in the local environment around the Hot DOGs. Considering that a high merger fraction has been found in a small sample of Hot DOGs based on HST/WFC3 imaging, we suggest that merger scenario could still demonstrate those results.

\begin{acknowledgements}

We thank the anonymous referee for constructive comments and suggestions.
We also thank Prof. Xu Kong, Prof. Hong-Xin Zhang, Mr. Zelin Xu, Mr. Mengqiu Huang, and Mr. Yongling Tang for valuable discussions. We acknowledge the support from National Key Research and Development Program of China (No. 2017YFA0402703). This work is supported by the National Natural Science Foundation of China (NSFC, grant Nos. 12173037, 11822303, 11773020, 12120101003, 11890691), and the China Manned Space Project with NO. CMS-CSST-2021-A04 and CMS-CSST-2021-A02. LF gratefully acknowledges the support of Cyrus Chun Ying Tang Foundations and the Strategic Priority Research Program of Chinese Academy of Sciences, Grant No. XDB 41010105.

\end{acknowledgements}

\end{CJK*}
\end{document}